\documentclass[12pt,a4paper, preprint, showkeys, superscriptaddress]{revtex4}
\usepackage[utf8]{inputenc}
\usepackage{hyperref}
\usepackage{pdflscape}
\usepackage{romannum}
 
 \usepackage{amsmath}
\usepackage{amssymb}
 \usepackage{subfigure}
 \usepackage{epsfig}
 \usepackage{epstopdf} 
\usepackage{amsmath,amsfonts}
\usepackage{graphicx}
\begin{document}
	\title{ Statistical description of galaxy clusters in Finzi model of gravity}
	\author {Abdul W. Khanday}
	\email{abdulwakeelkhanday@gmail.com}
	\affiliation{{Department of Physics, National Institute of Technology  Srinagar, Jammu and Kashmir -190006, India.}}
	\author {Sudhaker Upadhyay}
	\email{sudhakerupadhyay@gmail.com}
		\affiliation{Department of Physics, K. L. S. College,Magadh University Nawada, Bihar 805110, India}
	\affiliation{Inter-University Center for Astronomy and Astrophysics (IUCAA) Pune, Maharashtra-411007 }
 \affiliation{School of Physics, Damghan University, Damghan, 3671641167, Iran}
	\author {Naseer Iqbal}
\email{dni$_$phtr@kashmiruniversity.ac.in}
\affiliation{Department of Physics, University of Kashmir, Srinagar 190006, India}
\affiliation{Inter-University Center for Astronomy and Astrophysics (IUCAA) Pune, Maharashtra-411007 }

		\author {Prince A. Ganai}  
	\email{princeganai@nitsri.net}
\affiliation{{Department of Physics, National Institute of Technology  Srinagar, 
Jammu and Kashmir -190006, India.}}

\begin{abstract}
We exploit a new theory of gravity proposed by Finzi, which gives stronger interaction at large scales, to study the thermodynamic description of galaxy clusters. We employ  a statistical model to deduce various thermodynamics equations of state. In addition, we  analyze  the behavior of clustering parameter in comparison to its standard (Newtonian) counterpart.  The general distribution function and its behavior with varying strength   of clustering parameter are also studied. The possibility of phase transition is investigated and it is observed that a phase transition is  possible though hierarchically. We also analyze the model by comparing the results with data available through SDSS-\Romannum{3}, and obtain the parameters involved. 
	\end{abstract}	
	\maketitle 	
 \section{Introduction}
The distribution of matter on large scale is mainly described by the gravitational interaction. It is believed that the large scale structure of the universe is a gravitationally amplified descendant of a faint noise field believed to be seeded by quantum fluctuations in the early universe. A linear theory for   initial perturbations to the present observed matter distribution has been developed extensively~\cite{1}. 
The formation of first structures in the universe took place at a red-shift of $10-30$ in the dark matter halos of masses, $M> 10^5 - 10^8 M_\odot$~\cite{2}. This structure formation took place on the imprints of small matter density perturbations in the primordial matter density field. It has been verified by $N$-body simulations that the initial density perturbations have the potential to grow to the scale of  present day observed structure ~\cite{3}.
\par
It is well-known that the peculiar velocity of galaxies in a cluster doesn't agree well with the total mass of the visible matter. The estimated mass was  $200 - 400$ times less   than the mass required to prevent the rupture of the galaxies from the cluster ~\cite{4}. This led to the concept of dark matter. 
 There is a strong support to the dark matter hypothesis, but the identification of the particles that compose this matter is yet under discussion.   One explanation to this could be a strong correlation between the dark matter and  baryonic matter ~\cite{5}. The other intriguing idea could be a relevance of  modified Newtonian dynamics  (MoND) or any new dynamics on galactic scales.

The formation and distribution of galaxy clusters are crucial to understand the evolution of the structure formation in the universe.   
Saslaw and   Hamilton  
 developed a thermodynamic model for  nonlinear regime of the clustering of galaxies in an expanding universe \cite{7}. The theoretical predictions of this model involve the distribution of voids as well as galaxies corresponding to under-dense and over-dense regions, respectively, in the initial density field. The probability distribution function of $N$ galaxies found  in some volume $V$ predicted by this theory is given by 
\begin{equation}
f(N)=e^{-\bar{N}(1-b)-Nb}\frac{\bar{N}(1-b)}{N!}\left[\bar{N}(1-b)+Nb\right]^{N-1},\label{01}
\end{equation} 
where $\bar{N}$ is the ensemble-average  of particles found in any volume $V$ and $b$ is clustering parameter. 
\par
It is well known that the force accounting for the flat rotation curve of galaxies and  galaxy clusters requires stronger gravity at very larger spatial distances than produced by Newton's law of gravity.  
Recently, there have been several attempts  to account for this discrepancy through the modification to the general theory of relativity~\cite{8} .  For instance, in $f(R)$ group of theories,  a function $f(R)$ of Ricci scalar is used in place of Ricci scalar $R$ to account for the enhanced gravitational interaction on large scales.  
 The force field that governs the large scale structure formation also governs the distribution of mass at this scale. We can employ statistical methods to find the statistical distribution of matter at largest possible scales~\cite{9}. 
 Recently there has been  a lot of progress on the  study of the effect of modified  gravity laws  on
the clustering of galaxies using statistical mechanics \cite{10,11,12,13,14, 15,15a, 15b, 16,17,18, 19}. The effect of the modifications incorporated has been anticipated in the strength of the clustering parameter $b$.  For instance,  in~\cite{14}, the modification to the  clustering parameter as a function of correction factor has been studied.  
\par
 
In 1963, Finzi proposed a law of gravitation that gives a stronger interaction at relatively larger distances than predicted by Newton's law of gravity~\cite{20}. This force form can explain the larger velocities of galaxies in clusters without involving the concept of dark matter. The potential energy function proposed by  Finzi is given by 
\begin{equation}
	\Phi(r)=G_*m^2\left(\frac{1}{r^{1/2}}\right),\label{02}
 \end{equation}
where $G_*=-2k/\rho^{1/2}$, $k$ is a constant (equivalent to gravitational constant $G$) and $\rho$ is a characteristic length beyond which the potential becomes significant.
This potential   reduces to the usual Newtonian potential at   distances $\rho\approx r$. The correlation and distribution function of galactic clusters has not been studied for Finzi approximation. 
Here, we try to bridge this gap and also make a comparison of the effect of this model to the already studied theories. 

Our investigations are presented systematically as following. We construct the general partition function under Finzi potential for our system of galaxies in section \ref{sec2}.
Here, in order to avoid divergence, we consider extended nature of galaxies.
 We derive the various thermodynamic potentials along with their behavior under different circumstances,  e.g., Helmholtz free energy, pressure, internal energy, entropy and chemical potential in  section \ref{sec3}. 
In section \ref{sec4}, we compare the   clustering parameters with increasing radial distance corresponding to Newtonian and Finzi gravity. In section \ref{sec5}, we study the statistical distribution of galaxies in the new gravity law.The power-law behavior for the correlation function is presented in section \ref{sec6}. Finally we investigate the possibility of phase transition in section \ref{sec7}. We present the results  and their importance in the last section. 

\section{The partition function}\label{sec2}
In this  section, we develop the    general partition function by taking into consideration a system of gravitationally interacting particles, with the interaction defined by Finzi gravity. We assume the clustering of galaxies in the expanding background to be in 
quasi-equilibrium state  forming an ensemble of co-moving cells. We assume the system  consists of an ensemble of cells  of equal volume $V$ with number density $\bar{N}$. The form of the partition function for such a system of pairwise interacting particles having correlation energy $\Phi$ and average temperature $T$ is given by~\cite{9}  
\begin{equation}
 Z_{N}(T,V)=\frac{1}{N!}\left(\frac{2\pi mT}{\lambda^2}\right)^\frac{3N}{2} Q_{N}(T,V).\label{03}
\end{equation}
Here,$\lambda$ is the normalization constant and the factor  $N!$ takes into account the distinguishability of the system particles. The Boltzmann's  constant,$k_B$, is set equal to unity. Configuration part of the equation (\ref{03}) can be written as 
\begin{equation}
 Q_{N}(T,V)=\displaystyle \int....\int  \exp  \left[-T^{-1}\Phi(r_{1},r_{2},....r_{N})\right] d^{3N}r.\label{04}
\end{equation}
  The gravitational potential energy
function $\Phi(r_{1},r_{2},...r_{N})$ is a function of the relative positions  $r =|r_{i}-r_{j}|$ and is summed over all the pairs of particles. In the system of gravitationally interacting  bodies, the potential energy $\Phi(r_{1},r_{2},...r_{N})$ is due to all  pairs of particles present in   the system, i.e.
\begin{equation}
\Phi(r_{1},r_{2},...r_{N}) =\sum_{1\le i\le j\le N}^{}\Phi_{ij}(r).\label{05}
\end{equation}
With this simplification (\ref{05}), equation (\ref{04}) can now be written as 
 \begin{equation}
Q_{N}(T,V)=\displaystyle\int...\int\displaystyle \prod_{1\le i\le j\le N}\exp\left[-T^{-1}\Phi_{ij}(r)\right] d^{3N}r,\label{06}
\end{equation}
where $\Phi_{ij}$ is the  two point interaction energy between the $i^{th}$ and $j^{th}$ particle. The configuration integral can be written in terms of two-point function $f_{ij}(=e^{ -\frac{\Phi_{ij}}{T}}-1)$ defined as:
\begin{equation}
Q_{N}(T,V)=\displaystyle\int...\int(1+f_{12})(1+f_{13})(1+f_{23})(1+f_{14})...(1+f_{N-1,N})d^{3}r_{1}d^{3}r_{2}...d^{3}r_{N}.\label{08}
\end{equation}  
Here, we note that the  two-point function $f_{ij}$ is non-zero only when there are  interactions present. 
The Finzi potential (\ref{02}) diverges for the point-particle nature of galaxies. Therefore, 
we need to express potential (\ref{02})  by considering
the extended nature of galaxies (galaxies with halos).  This is done by introducing  the softening parameter $\epsilon$ $(0.01\le \epsilon \le 0.05)$ 
in the potential as
\begin{equation}
\Phi(r)= { \frac{G_*m^2}{(r +\epsilon)^{1/2}}}.\label{09}
\end{equation}
 
 By substituting the  two-point function corresponding to potential (\ref{09}) in equation (\ref{08}), the value of $Q_N$ for various values  of $N$ can be easily calculated. For instance, for $N = 1$, we have
\begin{equation}
Q_{1}(T,V)= V. 
\end{equation}
For $N=2$ value, we evaluate the integral (Eqn. \ref{08}) by fixing the  position of $r_1$ and integrating  over all the other system particles.  This simplifies the integral  to 
 \begin{equation}
Q_{2}(T,V)=V^{2}[1+X \alpha ],\notag
\end{equation}
where $ X=\frac{2G_*m^{2}}{Tr^{1/2}} $ and 
$\alpha=\sqrt{1+\frac{\epsilon}{r}}\big[3-4(\epsilon/r)+8 (\epsilon/r)^2-(\sqrt{\epsilon/r} 8(\epsilon/r)^2)\big]$.

Proceeding in  similar fashion , the values of $Q_N$ for $N = 3, 4, 5,6, 7,  ..., N$ can be easily calculated. For $N = 3$, we get
\begin{equation}
Q_{3}(T,V)=V^{3}\bigg[1+X \alpha \bigg]^{2}.\label{q_3}
\end{equation}
In general the above equation, Eqn.(\ref{q_3}), for $N$ number of  particles takes the form, 
\begin{equation}
Q_{N}(T,V)=V^{N}\bigg[1+X \alpha \bigg]^{N-1}.\label{11}
\end{equation}
Finally, substituting equation (\ref{11}) into equation (\ref{03}), we obtain the general partition function for an interacting  system of $N$ particles (galaxies)  as
\begin{equation}
Z_{N}(T,V)=\frac{1}{N!}\bigg(\frac{2\pi mT}{\Lambda^{2}}\bigg)^{\frac{3N}{2}} V^{N}\bigg[1+ \alpha X\bigg]^{N-1}.\label{12}
\end{equation}
Equation (\ref{12}) is the standard partition function (Canonical) of the system of $N$ particles interacting through the modified gravity law (\ref{02}). The correction to the partition function is inherent in the parameter $\alpha$.

\section{Thermodynamic equations of state of the system}\label{sec3}
The partition function equation (\ref{12}) contains all the necessary information about the macroscopic variable (free energy, entropy, internal energy, pressure, chemical potential ) of the system. The Helmholtz free  energy for the system of galaxies can be deduced from the partition function via the relation $F=-T\ln Z_{N}(T,V)$. For our system of galaxies the free energy  takes the following form: 
\begin{equation}
\begin{split}
F&=NT\ln\bigg(\frac{N}{V}T^{\frac{-3}{2}}\bigg)-NT-\frac{3}{2}NT\ln\bigg(\frac{2\pi mT}{\lambda^{2}}\bigg)-NT\ln [1+X \alpha ],\\
&=NT\ln\bigg(\frac{N}{V}T^{\frac{-3}{2}}\bigg)-NT-\frac{3}{2}NT\ln\bigg(\frac{2\pi mT}{\lambda^{2}}\bigg)+NT\ln [1-b_n].
\label{13}
\end{split}
\end{equation}
Here, we approximated $N-1\approx N$. In equation (\ref{13}), the parameter $b_n$ is defined as
\begin{equation}
b_n=\frac{\alpha X}{ 1+\alpha X }. \label{14}
\end{equation} 
 This is the modified clustering parameter which contains information of the strength of correlation that governs the time evolution of the galaxy cluster and can take values between $0$ and $1$.

\begin{figure}[h!]
	\centering
	\includegraphics[width=8 cm, height=6 cm]{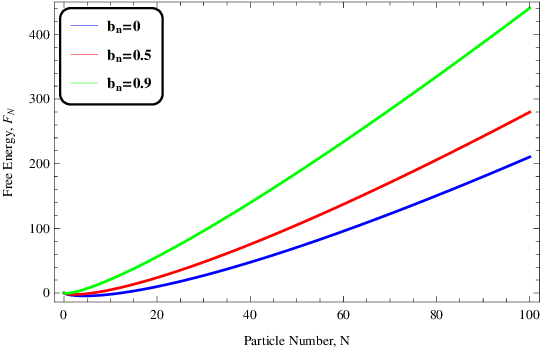}
	\caption{Variation of the free energy of the system  with  particle number  for various values of the $\alpha$ and unit value for the rest of parameters.}\label{fig1}
\end{figure}

 Once the  free energy is known,  other  thermodynamic equations of state  can be estimated easily. For example, we can calculate  the entropy of the system of galaxies utilizing the relation, $S=-\bigg(\frac{\partial F}{\partial T}\bigg)_{V,N}$. Substituting equation(\ref{13}) in this relation the entropy of the system takes the following form:
\begin{equation}
S=N\ln\biggl(\frac{V}{N}T^{3/2}\biggl) +N \ln [1+X \alpha  ] -3N\frac{X \alpha }{ 1+X\alpha } + \frac{5}{2}N +\frac{3}{2}N \ln\bigg(\frac{2\pi m }{\lambda^{2}}\bigg).\label{15}
\end{equation}
\par
Specific entropy i.e., entropy per particle of the system corresponding to equation (\ref{13}) takes the following form

\begin{equation}
\frac{S}{N}=\ln\biggl(\frac{V}{N}T^{3/2}\biggl)-\ln[1-b_n]-3b_n
+\frac{5}{2}+\frac{3}{2}\ln \frac{2\pi m}{\lambda^2},\label{16}
 \end{equation}
\begin{figure}[h!]
	\centering
	\includegraphics[width=8 cm, height=6 cm]{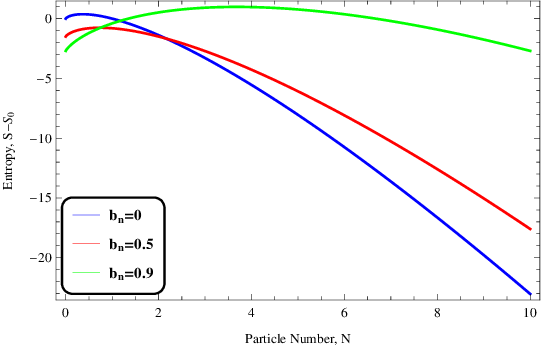}
	\caption{ The  variation of entropy ($S-S_0$) with an increasing particle number in the system  for various values of clustering parameter $b_n$.}\label{fig2}
\end{figure}

The total internal energy of the interacting  system of galaxies can be calculated using the basic definition, $ U = F + T S$. Upon substituting the values for free energy ($F$) (\ref{13})  and entropy ($S$)  (\ref{14}), the relation for internal energy of the system in terms of the new clustering parameter takes the form
 \begin{eqnarray}
U&=&\frac{3}{2}NT\bigg[1-2\frac{X\alpha  }{1+X\alpha }\bigg]\nonumber\\
&=&\frac{3}{2}NT\left[1-2b_n\right].\label{17}
\end{eqnarray}
 \begin{figure}[h!]
	\centering
	\includegraphics[width=8 cm, height=6 cm]{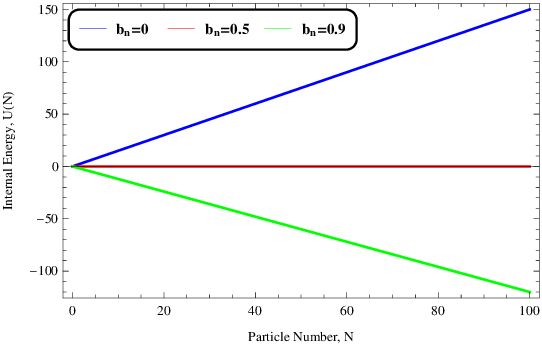}
	\caption{ The  variation of internal energy function $U$ of the system with increasing  number of particles for various values of correlation parameter $b_n$.}\label{fig3}
\end{figure}
Figure (\ref{fig3}) shows the graphical visualization of the effect of modified clustering parameter $b_n$ on the internal energy function of the system of galaxies interacting gravitationally. \\
The equation of the pressure caused by the particles in the system can be obtained utilizing the fundamental relations $P=-\bigg(\frac{\partial F}{\partial V}\bigg)_{T,N}$. Using relation (\ref{13}), the pressure of the system takes the form:

\begin{align}
P&=\frac{NT}{V} \left[1-\frac{X \alpha }{1+X \alpha } \right],
\nonumber\\ 
&=\frac{NT}{V}\left[1-b_n\right].\label{18}
\end{align}

\begin{figure}[h!]
	\centering
	\includegraphics[width=8 cm, height=6 cm]{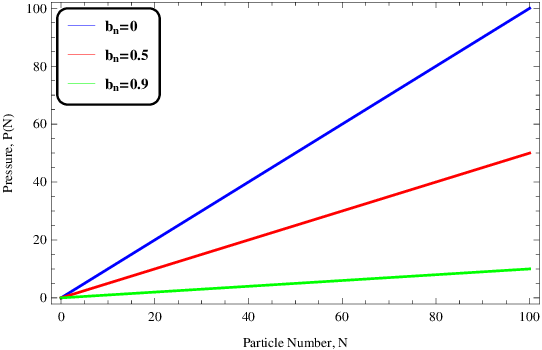}
	\caption{ The  variation of the pressure with increasing particle number  for various values of clustering parameter $b_n$.}\label{fig4}
\end{figure}
\par
\begin{figure}[h!]
	\centering
	\includegraphics[width=8 cm, height=6 cm]{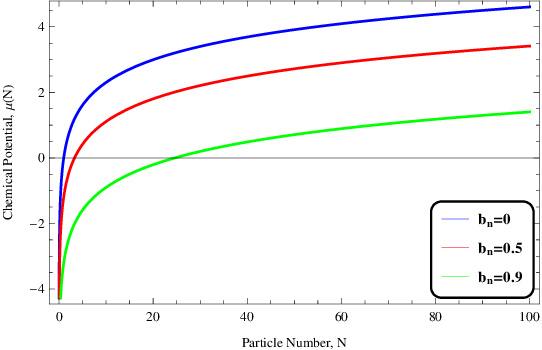}
	\caption{ The graph visualizes the behavior of  the chemical potential $\mu (N)$ of the system for various values of the modified clustering parameter $b_n$.} \label{fig5}
\end{figure}

 
Finally,  we  deduce the relation for the chemical potential $\mu$ of the system using the fundamental relation,
 $\mu=\bigg(\frac{\partial F}{\partial N}\bigg)_{T,V}$ as
\begin{eqnarray}
\mu &=&T\bigg(\ln\frac{N}{V}T^{-\frac{3}{2}}\bigg)+T\ln\bigg[1-\frac{X  }{1+X  }\bigg]-T\frac{X }{1+X 
 }-\frac{3}{2}T\ln\bigg(\frac{2\pi m}{\lambda^{2}}\bigg),\nonumber\\
&=&T\left(\ln \frac{N}{V}T^{-3/2}\right)+T\ln \left[1-b_n\right]-Tb_n
-\frac{3}{2}T\ln\left(\frac{2\pi M}{\lambda^2}\right).\label{19}
\end{eqnarray} 
The graphical visualization  of the variation  in the chemical potential of the system  with an increase in the particle number can be seen in figure (\ref{fig5})
 \section{A Comparison of New and old clustering parameter}\label{sec4}
 In this section the clustering parameter $b_n$ developed in this work based on the Finzi model of gravity is compared with the standard parameter $b$ defined in~\cite{9}. The correlation parameter gives the strength of correlation and will in turn decide the time-length of clustering.
 \begin{figure}[h!]
 	\centering
 	\includegraphics[width=8 cm, height=7 cm]{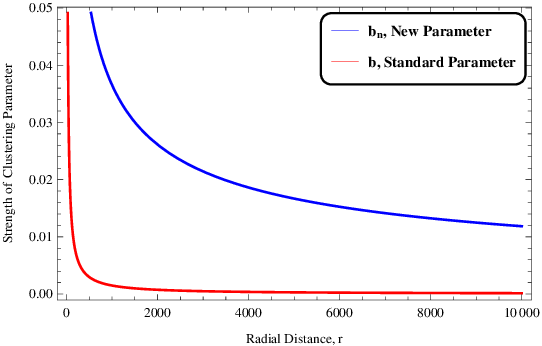}
 	\caption{Comparison of the standard correlation parameter $b$ and modified clustering parameter $b_n$}\label{fig6}
 \end{figure}
 From the graph (Fig.(\ref{fig6})) it can been seen that the new clustering parameter is more strong than the standard one. This is due to the effect of the increased strength of the potential energy function at large distances as defined in the Finzi model. 
\section{General form of the distribution function}\label{sec5}
\begin{figure}[h!]
	\centering
	\includegraphics[width=8 cm, height=6 cm]{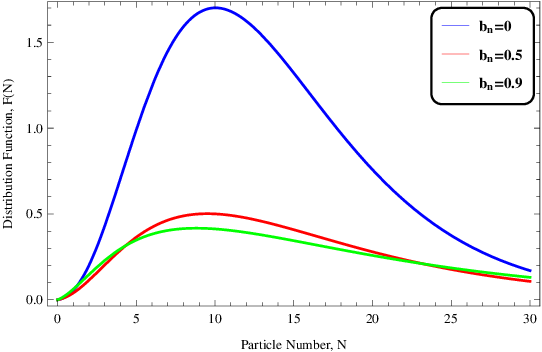}
	\caption{ Graphical representation of the distribution function $F(N)$ with a change in the  number of particle in the system, $N$ for various values of $b_n$.}
	\label{fig7}\end{figure}

The general form of the distribution function $f(N)$,  which characterizes the galaxy clustering, describes the distribution of voids as well as the number of galaxies in  fixed volume cells distributed through out
the system. Here we let the system particles cross the cell boundaries which in turn can change the particle number in each cell. Thus we derive  the  grand canonical partition function defined as  
\begin{equation}
	Z_G(T,V,z)=\sum_{N=0}^{\infty}\exp \left(\frac{N\mu}{T}Z_N(T,V)\right).\label{20}
\end{equation} 
 The probability distribution function   of  $N$ particles contained in cells  of fixed volume $V$ in a  grand canonical ensemble is given by
\begin{eqnarray}
F(N)&=&\sum_{i=0}^{N}\frac{\exp\frac{N\mu}{T}\exp\frac{-U}{T}}{Z_G(T,V,z)},\nonumber\\
&=&\frac{\exp\frac{N\mu}{T}Z_N(T,V)}{Z_G(T,V,z)}.\label{21}
\end{eqnarray} 
The factor $z=\exp\frac{\mu}{T}$ is the fugacity of the system and it determines the activity within the system. From this basic  equation (\ref{21}), the general form of the distribution function of  system can be determined easily. Utilizing the partition function (\ref{20}) with chemical potential  equation (\ref{12})  the distribution function of the system takes the  following form:
\begin{equation}
F(N)= \frac{\bar{N}}{N!}(1-b_n)\bigg[\bar{N}(1-b_n)+Nb_n\bigg]^{N-1} \exp-Nb_n-\bar{N}(1-b_n).\label{22}
\end{equation}
From equation (\ref{22}) we confirm that there is no change in the  basic structure of the distribution function in Finzi gravity model  to the one derived by Saslaw and Hamilton for Newtonian gravity~\cite{7, 13}. The graphical representation of the probability distribution function with increasing particle number $N$  is shown in figure \ref{fig7}. From the figure, it is clearly visible that the peak value in the  distribution function decreases with an increase in the  value of the clustering parameter.
\section{Power law for two-point correlation function}\label{sec6}
Here we study the behavior  of  the two-point correlation function in the new model of gravity. 
We write the clustering parameter in the form~\cite{21}
\begin{equation}
b_n=\frac{G_* \bar{\rho}}{6T}\int \left[\frac{1}{(r+\epsilon)^{1/2}}\right]\xi_2(\rho,r,T) d{V}, \label{23}
\end{equation}
where $\bar{\rho}$ is the number density and $\xi_2$ is the two-point correlation function. 
\par
Differentiating with respect to $V$, equation (\ref{23}) yields
\begin{equation}
\begin{split}
\frac{\partial b_g }{\partial V}  =\frac{G_* \bar{\rho}}{6T}\frac{\partial}{\partial V} \int \left[\frac{1}{(r+\epsilon)^{1/2}}\right]\xi_2(\rho,r,T) d{V}\\
+\frac{G_*}{6T}\left(-\frac{\partial \bar{\rho}}{\partial V} \right)\int \left[\frac{1}{(r+\epsilon)^{1/2}}\right]\xi_2(\rho,r,T) d{V},\label{24}
\end{split}
\end{equation}
where we have used $\frac{\partial V}{\partial{ \bar{\rho}}} {\bar{\rho}}=-\frac{\bar{\rho}}{V}$. 

\begin{figure}[h!]
  	\centering
  	\includegraphics[width=8 cm, height=6 cm]{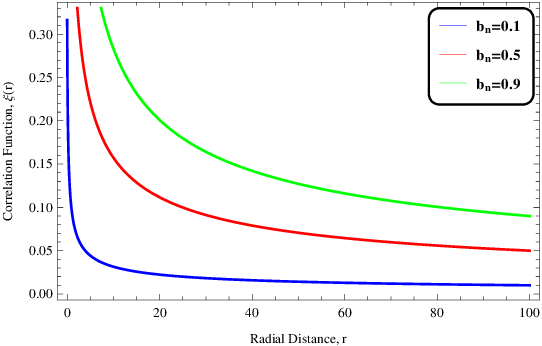}
  	\caption{ Graphical representation of the two point correlation function $\xi(r)$ with increasing radial distance, $r$, for different values of clustering parameter $b_n$.}
  	\label{fig8}
  \end{figure}

Using the relation $\frac{\partial b_n } {\partial\bar{\rho}}=\frac{b_n(1-b_n)}{\bar{\rho}}$ and equation (\ref{24}), we obtain the following expression for the power law
\begin{equation}
\xi_2(r)=\frac{9Tb_n^2}{2\pi G_* \bar{\rho} }\left(\frac{1}{(r+\epsilon)^{1/2}}\right).\label{25}
\end{equation}
The graphical representation of the two point correlation function can be visualized from the figure (\ref{fig8}). We can see the strength of the correlation function decreases with increasing radius. This leads us to an important result about the correlation of system points in different cells.The galaxies within a cell are more correlated than the galaxies in the adjacent cells. Thus the transfer of system points from cell to cell are less likely although not negligible. 
\section{Possibility of phase transition }\label{sec7}
If an interaction is introduced in a system characterized by Poisson distribution the system has a high chance of changing phase from less correlated to highly correlated. The possibility of phase transition in case of our system of galaxies interacting gravitationally can not be ignored.  Here we will try to find out if the Finzi interaction can cause a phase transition in the system of gravitationally interacting point particles. The result is important as the possible phase transition can break the homogeneity of the system and cause lumpiness in the structure. Among many indicators of phase transition, specific heat is an important candidate to track.

\begin{figure}[h!]
	\centering
	\includegraphics[width=8 cm, height=6 cm]{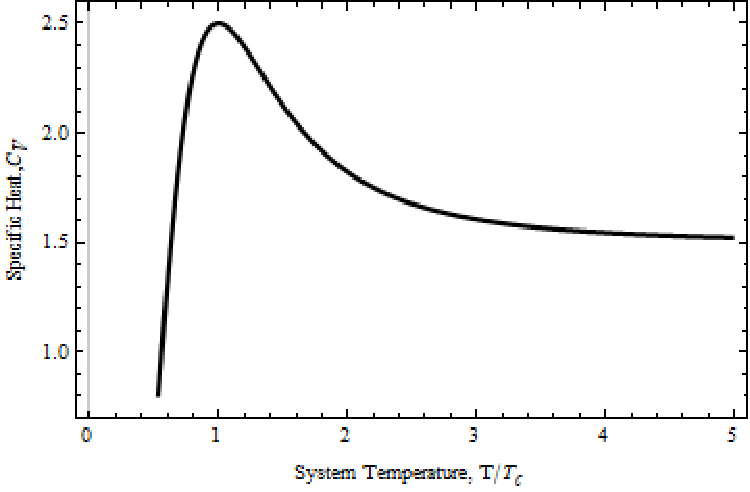}
	\caption{The behavior of specific heat $C_V$ with changing system temperature  $T/T_C$. We observe that the system symmetry breaks around the critical temperature, $T=T_C$. }\label{fig9}
\end{figure}

 The specific heat (at constant volume) $C_V$ is defined as
\begin{equation}
	C_V=\frac{1}{N}\left(\frac{\partial U}{\partial T}\right)_{N,V}. 
\end{equation}
Using the relation for internal energy (\ref{17}), the specific heat of the system takes the following form
\begin{equation}
	C_V=\frac{3}{2}\left[\frac{1+6\alpha  X -4\alpha^2 X^2}{\left(1+\alpha X\right)^2}\right].\label{27}
\end{equation} 
As $b_n\rightarrow 0, C_V\rightarrow 3/2$, this corresponds to no interaction among the system particles. As $b_n\rightarrow 1, C_V\rightarrow -3/2$, which means the system is  fully virialized. Between these two extreme values lie the maximum values of specific heat at some critical value of temperature. This extreme value of specific heat indicates a possible phase transition at $T=T_C$. 
\begin{equation*}
	\frac{\partial C_V}{\partial T}\biggr\rvert_{T=T_C}=0.
\end{equation*} 
     This gives an expression for the critical temperature as
\begin{equation}
	T_C=\left[3\frac{\bar{N}}{V}\left(GM^2\right)^3 \alpha \right]^{1/3}.\label{28}
\end{equation}
 In terms of critical temperature the specific heat of the system, $C_V$,  given in (\ref{27})  can be written as 
\begin{equation}
 C_V=\frac{3}{2}\left[1-2\frac{1-4\left(T/T_C\right)^3}{\left\{1+2\left(T/T_C\right)^3\right\}^2}\right].
\end{equation}
At $T=T_C, C_V=5/2$, a property of a diatomic gas. This is an indication of   system symmetry breaking  due to the formation of binaries. The phase transition is hierarchical and not spontaneous as observed in many other physical systems. The graphical behavior of the specific heat is visualized in figure (\ref{fig9}).   
\section{observational data}
In this section we test our model with the data  obtained through Sloan Digital Sky Survey \Romannum{3}  (SDSS-\Romannum{3}) through its newest Data Release (DR12). SDSS-\Romannum{3} contains additional sky coverage and better galaxy estimates  than SDSS-\Romannum{1} and SDSS-\Romannum{2}. The data is present in catalog \cite{22}. In figure (\ref{fig11}) the sky distribution of galaxy clusters in various  RA and DEC coordinates is presented from this catalog. This catalog contains an information (RA, DEC, z, N etc) of almost 132,684 clusters. The catalog gives a parameter $r_{200}$ which is the distance up to which mean density is $\approx$ 200 and also the number of galaxy clusters in it i,e $N_{200}$. Figure (\ref{fig10},(a)-(i)) shows the model fitted to the data.

First we bin the data  on the basis of radius $(R)$ and Redshift $(z)$. The bin size is chosen to be  $\Delta R\approx 0.35 Mpc$ and $\Delta z\approx 20$. We divide cells by physical boundaries i.e., $R$. The probability distribution of each cluster  is determined by substituting arbitrary values for the fitting parameter $b_n$ and $\rho$. The model is then fit through Application Programming Interface (API) Scipy.optimize.curve\_fit of SciPy python Library. We obtained the optimized values for the parameters involved in equation(\ref{22}), listed in table(\ref{table}).
\par
The value of the charaecteristic length, $\rho$, is obtained through the relation,
\begin{equation}
\sqrt{\rho}= \frac{1-b_n}{b_n}.\label{29}
\end{equation}
We have set all other parameters to unity in (\ref{29}). The characteristic length $\rho$ is measured in the same units as $r$. For instance, the value $\rho = 5$ means for 1 unit of $r$, the characteristic length is 5 units.
	\begin{figure}[h!]
	\centering
	\subfigure[]{\includegraphics[width=0.40\textwidth]{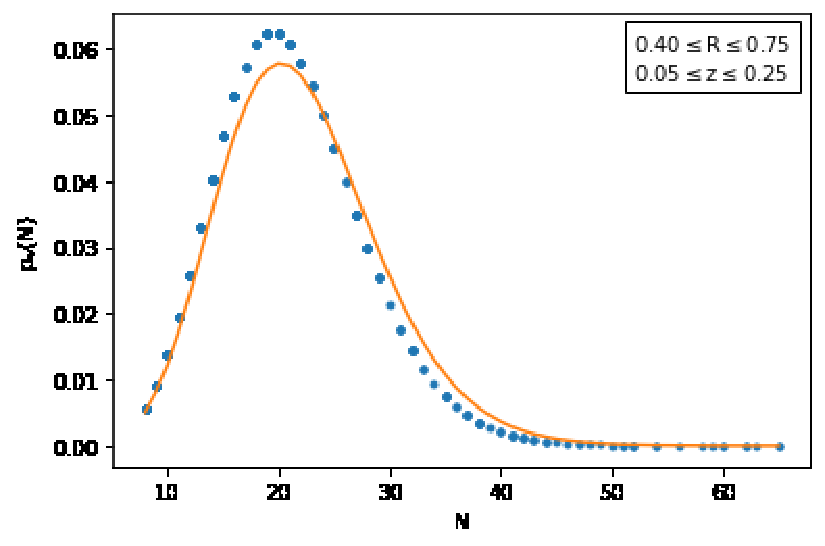}} 
	\subfigure[]{\includegraphics[width=0.40\textwidth]{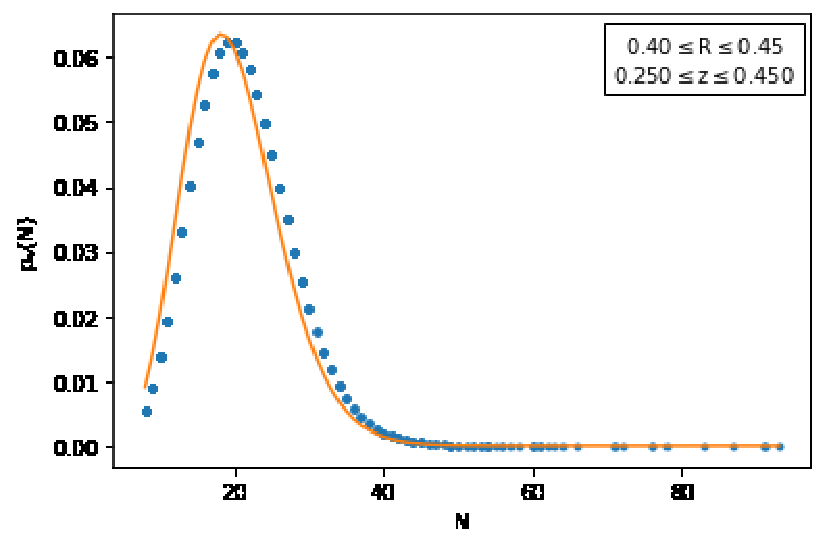}}\\
	\subfigure[]{\includegraphics[width=0.40\textwidth]{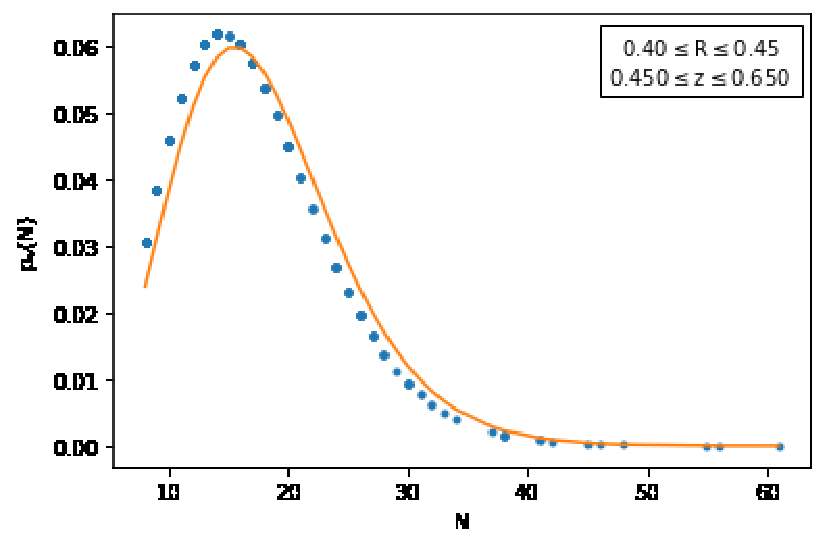}}
	\subfigure[]{\includegraphics[width=0.40\textwidth]{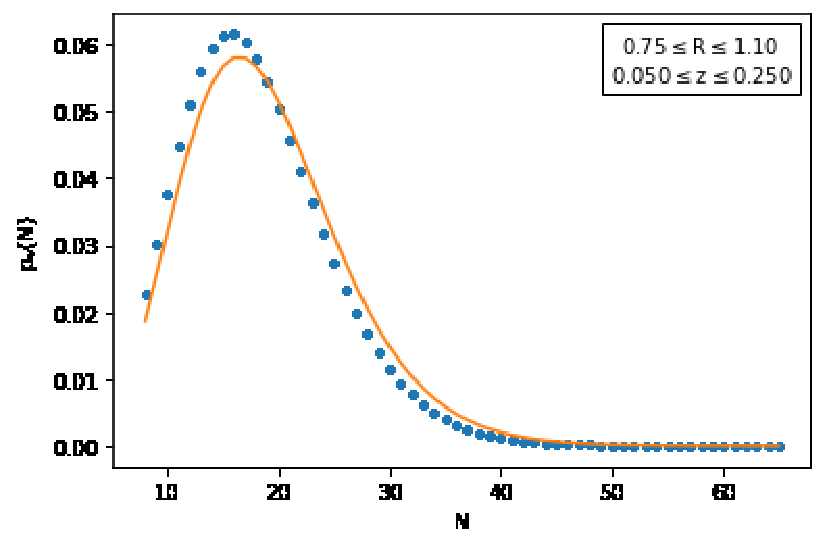}}\\
	\subfigure[]{\includegraphics[width=0.40\textwidth]{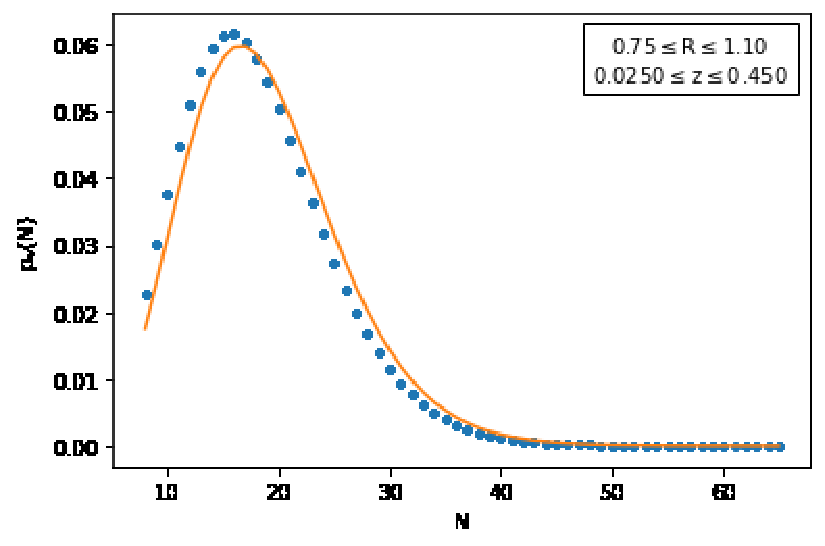}} 
	\subfigure[]{\includegraphics[width=0.40\textwidth]{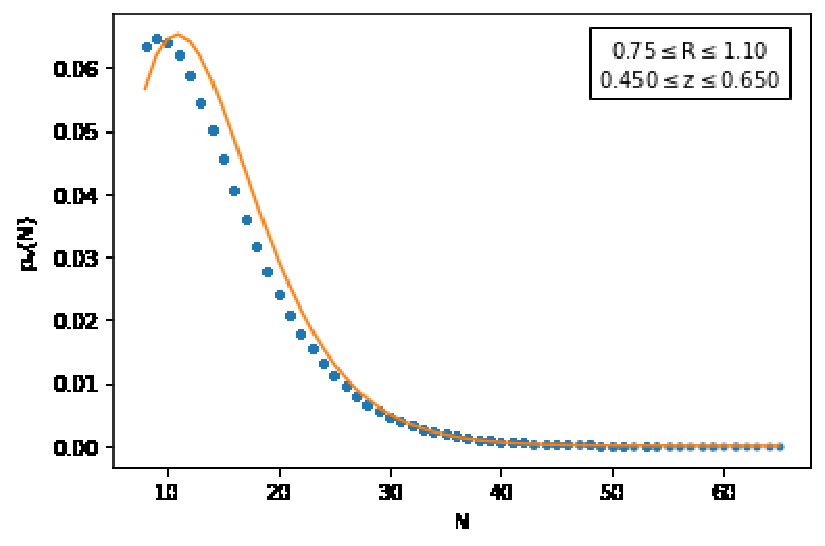}}\\
\end{figure}
\begin{figure}
	\subfigure[]{\includegraphics[width=0.40\textwidth]{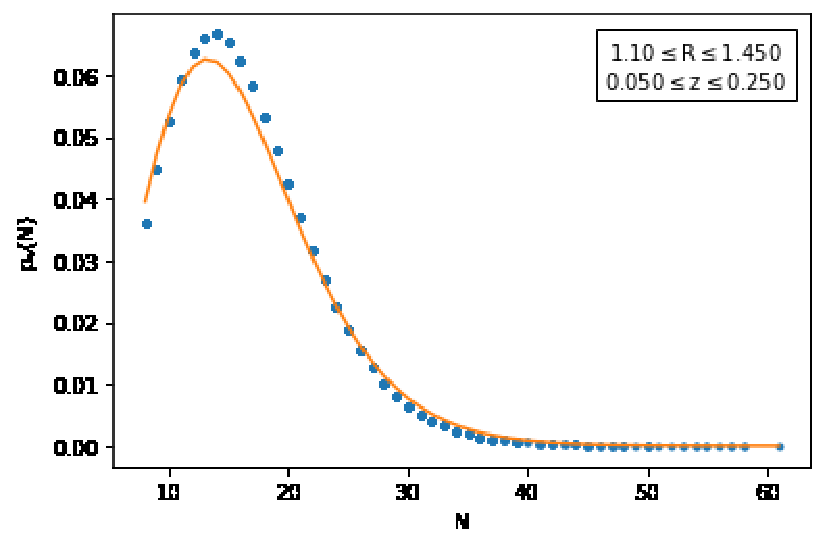}}
	\subfigure[]{\includegraphics[width=0.40\textwidth]{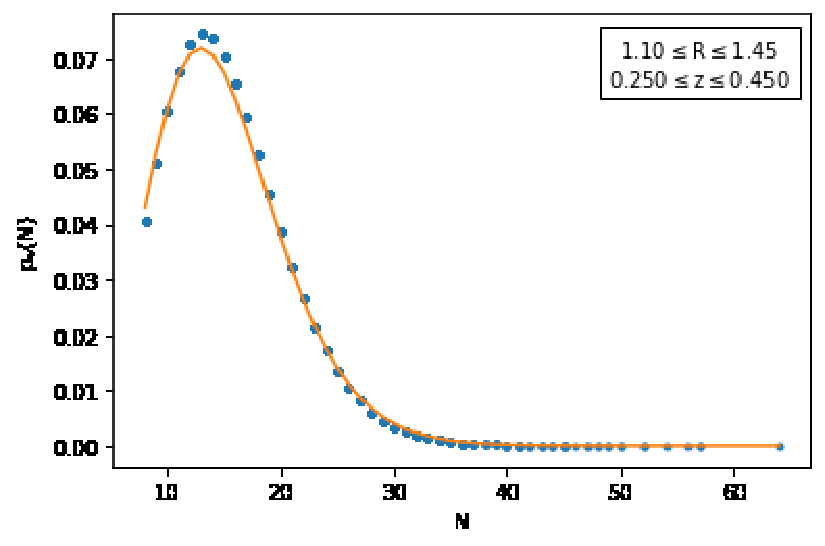}}\\
	\subfigure[]{\includegraphics[width=0.40\textwidth]{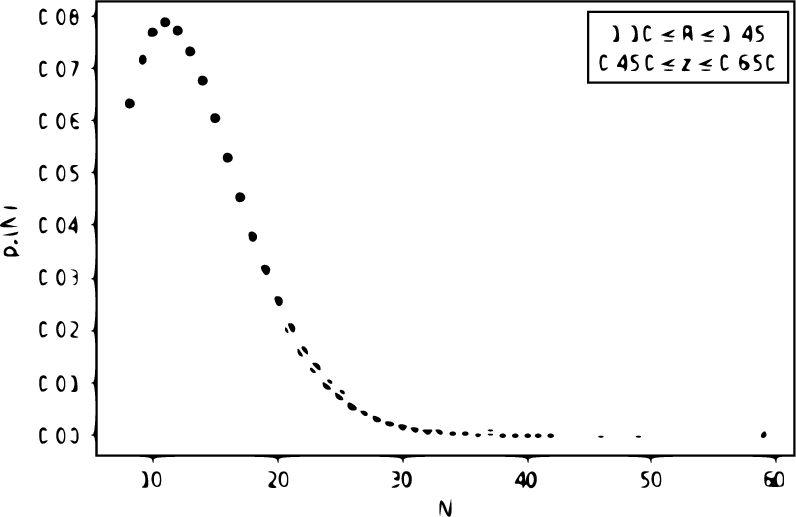}} 
	
	\caption{The theoretical (solid line) and calculated (dotted line) probability distribution of galaxy clusters in various red-shift and radius bins}
	\label{fig10}
\end{figure}
\begin{figure}
	\centering
	\subfigure[]{\includegraphics[width=0.40\textwidth]{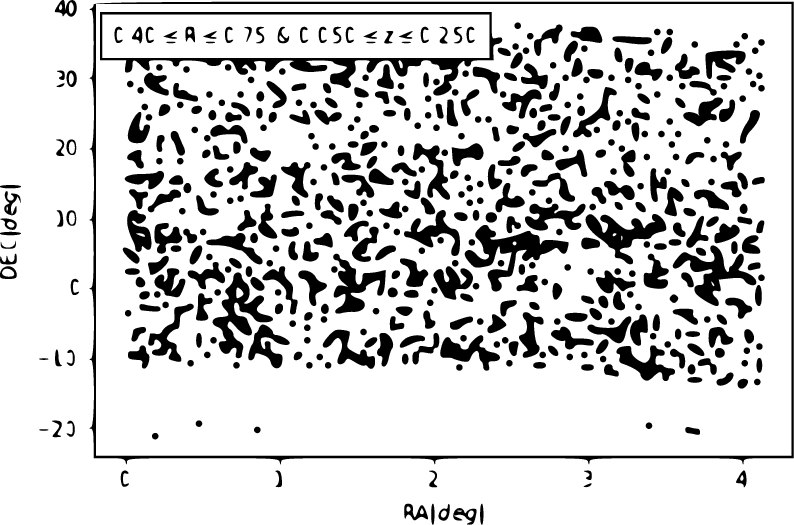}} 
	\subfigure[]{\includegraphics[width=0.40\textwidth]{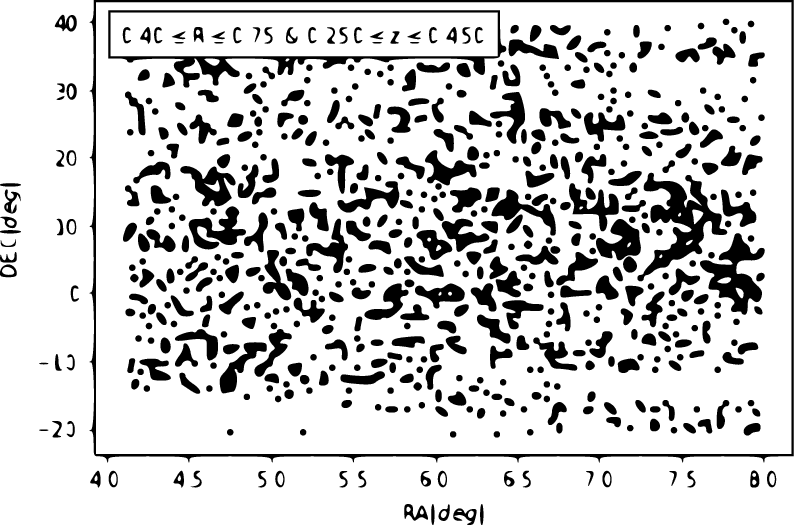}} \\
	\subfigure[]{\includegraphics[width=0.40\textwidth]{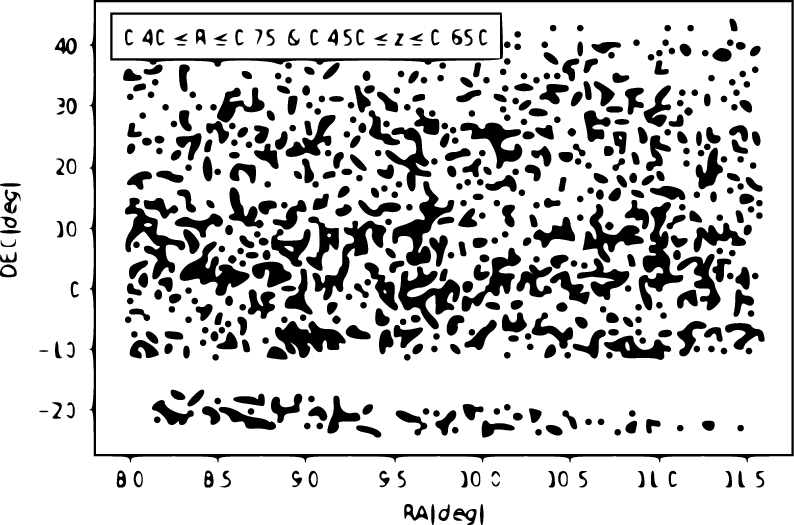}}
	\subfigure[]{\includegraphics[width=0.40\textwidth]{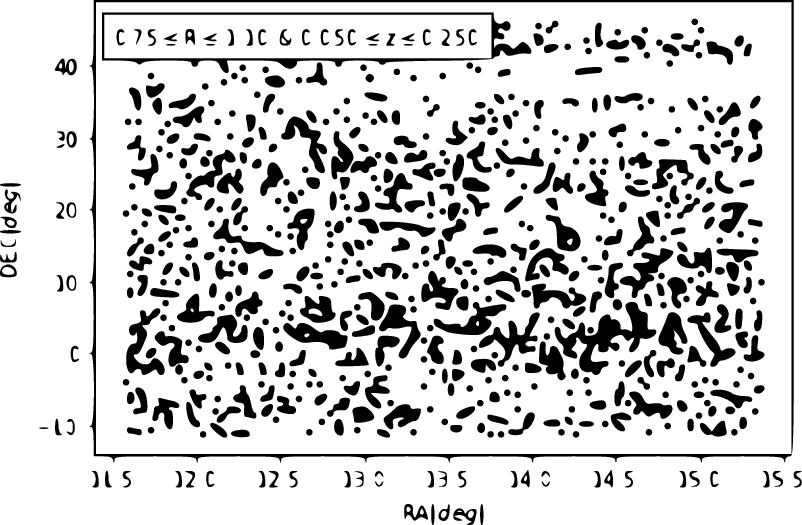}}\\
	\subfigure[]{\includegraphics[width=0.40\textwidth]{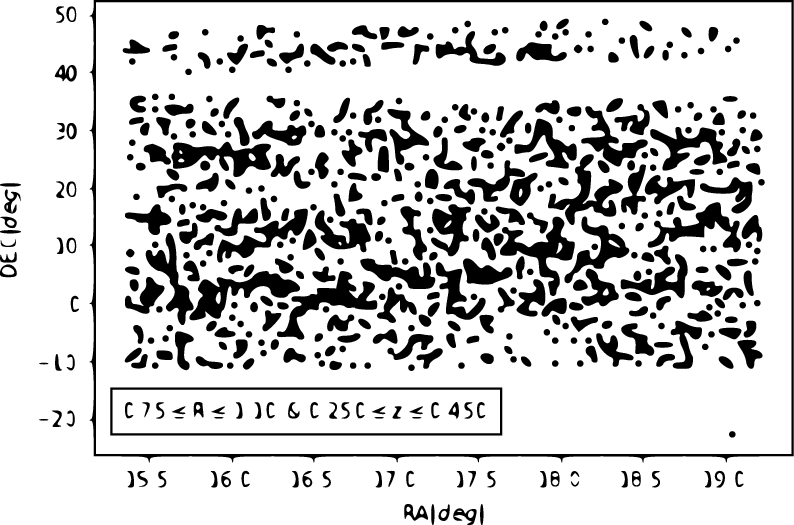}} 
	\subfigure[]{\includegraphics[width=0.40\textwidth]{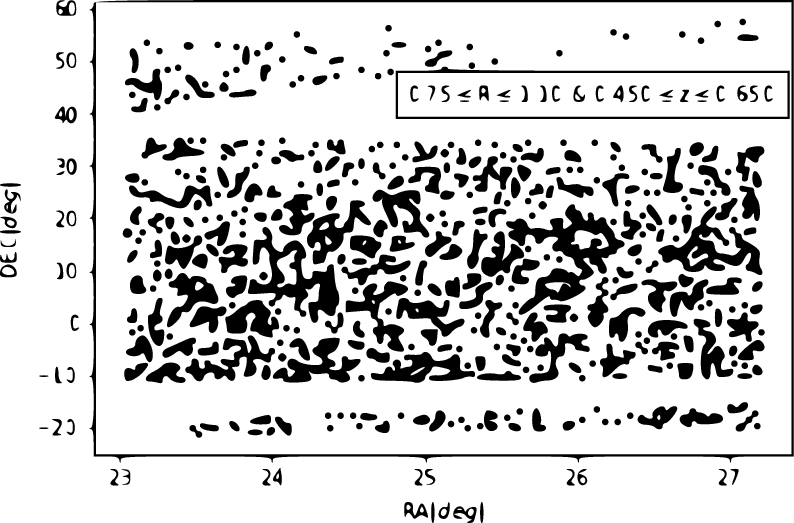}} 
\end{figure}
\begin{figure}
	\subfigure[]{\includegraphics[width=0.40\textwidth]{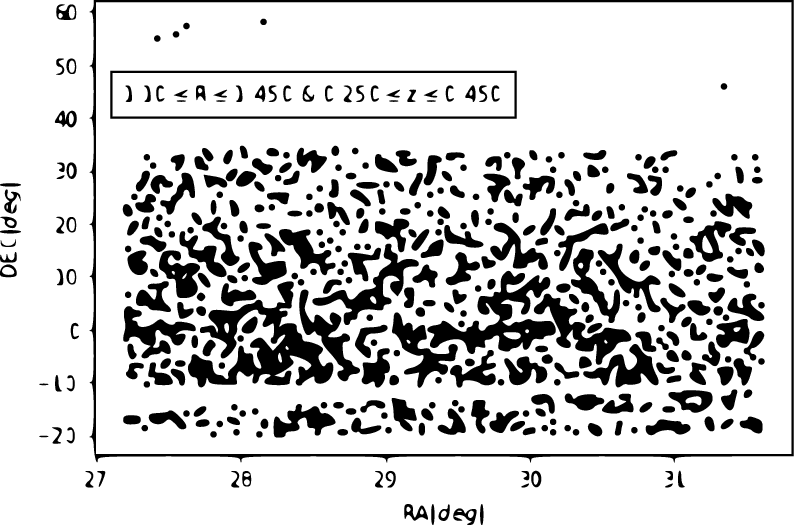}}
	\subfigure[]{\includegraphics[width=0.40\textwidth]{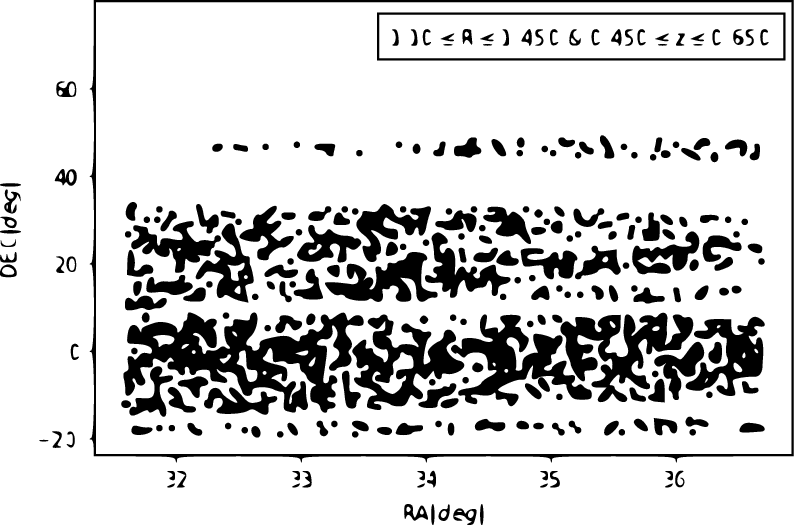}}\\
	\subfigure[]{\includegraphics[width=0.40\textwidth]{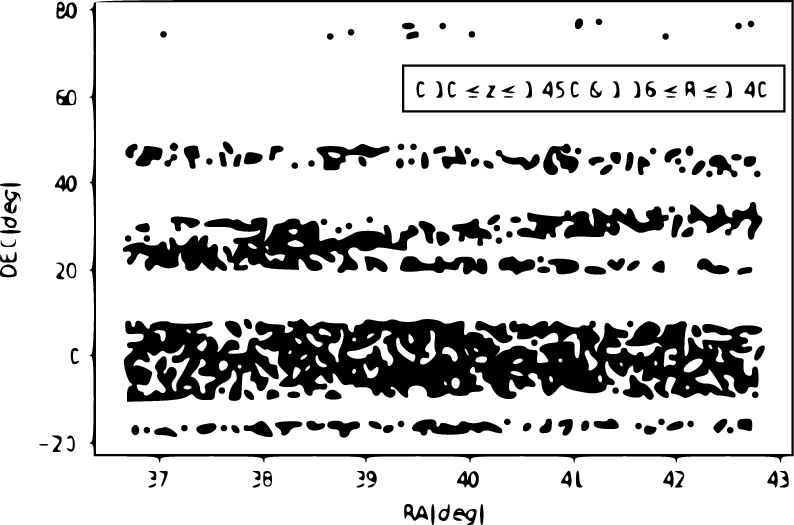}} 
	
	\caption{The figure  shows the sky distribution of galaxies  in different RA and DEC coordinates (measured in degrees)}
	\label{fig11}
\end{figure}
\begin{table*}
	\begin{minipage}{\textwidth}
		\caption{ The table shows the various parameters (given/fitting) after fitting the model to data points. The data points are divided first in radius bins and then in Redshift bins. $N_t$ is the number of clusters in each bin. }\label{tab:results}
		\centering
		\resizebox*{\textwidth}{!}{
			\begin{tabular}{|c|cccc|cccc|cccc|}
				\multicolumn{13}{c}{} \\
				\hline
				& \multicolumn{4}{c|}{$0.40<R<0.75 Mpc$} & \multicolumn{4}{c|}{$0.75<R<1.10 Mpc$} & \multicolumn{4}{c|}{$1.10<R<1.45 Mpc$} \\
				 $z$          & $N_G$ & $\bar{N}$                & $b_n$           & $\rho$ & $N_{G}$ & $\bar{N}$ & $b_n$ & $\rho$ 
				& $N_{G}$ & $\bar{N}$ & $b_n$ & $\rho$   \\
				\hline
				$[0.05,0.250]$ &  $8259$ & $21.123$ & $0.30$ & $5.44$ & $368936$ &
				$16.876$ & $0.38$ & $2.66$ & $91826$ & $15.498$ & $0.0.37$ &
				$2.89$  \\
				$[0.251,0.450]$ &  $19488$ & $20.406$ & $0.35$ & $3.45$ & $789028$ &
				$15.227$ & $0.31$ & $4.95$ & $146297$ & $14.591$ & $0.32$ &
				$4.51$  \\
				$[0.450,0.650]$ &  $4995$ & $16.594$ & $0.40$ & $2.25$ & $387334$ &
				$12.536$ & $0.50$ & $1$ & $62609$ & $12.317$ & $0.33$ &
				$4.12$ \\	
				\hline	
		\end{tabular}}\label{table}
	\end{minipage}
\end{table*}

From the plots (\ref{fig10}), we observe that the model fits very closely to data in redshift ranges $0.250<z<0.450$ and radius ranges $0.40<R<0.45,1.0<R1.45$, fig(\ref{fig10}(b,e)). In bins $0.450<z<0.650$;$0.75<R<1.10$, (fig\ref{fig10} (f)) the model does not fit very closely to the data.

\section{Conclusion}\label{sec8}
The velocities of galaxies in a cluster are much higher than depicted by the visible matter which brings in the concept of gravitating non-visible matter called dark matter. While the hunt for the exact description is underway, it is plausible to reconsider Newtonian dynamics and modify it to fit the observed data. 

In this work, we have considered a  gravitational potential proposed by   Finzi and,  by using  statistical methods, we deduced various thermodynamic quantities along with correlation function. A graphical analysis of  various thermodynamic equations of state was also done.  A new clustering parameter was deduced based on this theory which shows a greater strength as compared to the standard one.  We have also made a graphical comparison of the the clustering parameters for Finzi and Newtonian gravity. 
From the behavior of the clustering parameter, it is obvious that the Finzi model of gravity at larger scale is more significant.  The power-law of the correlation  function for Finzi model is also discussed. 
 We also studied  possibility of phase transition within the system. The comparison of our results with the data was also studied and it could be seen that the model fits the data very closely in some Redshift $z$ and radius ranges $R$ , while in some regions the fit is not too appropriate although there is agreement with the trend.

  \end{document}